\documentclass[letterpaper,twocolumn,10pt]{article}

\usepackage{paralist}
\usepackage{xurl}
\usepackage{usenix-2020-09}
\usepackage{amsmath}
\usepackage{graphicx}
\usepackage{tabularx}
\usepackage{booktabs}
\usepackage{colortbl}
\usepackage{xcolor}
\usepackage{multirow}
\usepackage{enumitem}
\usepackage[affil-it]{authblk}
\usepackage{nopageno}
\usepackage{dblfloatfix}
\usepackage[available,functional,reproduced]{usenixbadges}

  {\begin{itemize}[leftmargin=*]%
    \setlength{\itemsep}{0pt}%
    \setlength{\parskip}{0pt}%
    \setlength{\parsep}{0pt}}%
  {\end{itemize}}

  {\begin{enumerate}[leftmargin=*]%
    \setlength{\itemsep}{0pt}%
    \setlength{\parskip}{0pt}%
    \setlength{\parsep}{0pt}}%
  {\end{enumerate}}

\newcommand{\name}{\textsc{Amides}}

\definecolor{Gray}{gray}{0.9}

\makeatletter
\renewcommand\AB@affilsepx{\qquad \protect\Affilfont}
\makeatother

\begin{document}

\date{}

\title{\Large \bf You Cannot Escape Me: Detecting Evasions of SIEM Rules in Enterprise Networks}

\author[*]{Rafael Uetz}
\author[*]{Marco Herzog}
\author[*]{Louis Hackländer}
\author[$\dagger$]{Simon Schwarz}
\author[$\ddagger$,*]{Martin Henze}

\affil[*]{Fraunhofer FKIE}
\affil[$\dagger$]{University of Göttingen}
\affil[$\ddagger$]{RWTH Aachen University}

\maketitle

\begin{abstract}
Cyberattacks have grown into a major risk for organizations, with common consequences being data theft, sabotage, and extortion.
Since preventive measures do not suffice to repel attacks, timely detection of successful intruders is crucial to stop them from reaching their final goals.
For this purpose, many organizations utilize Security Information and Event Management (SIEM) systems to centrally collect security-related events and scan them for attack indicators using expert-written detection rules.
However, as we show by analyzing a set of widespread SIEM detection rules, adversaries can evade almost half of them easily, allowing them to perform common malicious actions within an enterprise network without being detected.
To remedy these critical detection blind spots, we propose the idea of adaptive misuse detection, which utilizes machine learning to compare incoming events to SIEM rules on the one hand and known-benign events on the other hand to discover successful evasions.
Based on this idea, we present \name{}, an open-source proof-of-concept adaptive misuse detection system.
Using four weeks of SIEM events from a large enterprise network and more than 500 hand-crafted evasions, we show that \name{} successfully detects a majority of these evasions without any false alerts.
In addition, \name{} eases alert analysis by assessing which rules were evaded.
Its computational efficiency qualifies \name{} for real-world operation and hence enables organizations to significantly reduce detection blind spots with moderate effort.
\end{abstract}

\section{Introduction}
\label{sec:introduction}

Enterprise networks increasingly face severe cyberattacks~\cite{hiscox2022cyber}, e.g., illustrated by the 2020 SolarWinds hack affecting thousands of companies and government agencies~\cite{reuters2021solarwinds}.
Besides considerable efforts to roll out preventive measures, enterprises additionally depend on threat detection to uncover successful adversaries breaching their line of defense~\cite{lord2022investments}.
For this purpose, many enterprises utilize \emph{Security Information and Event Management} (SIEM) systems to collect and automatically analyze large numbers of security-related events (e.g., user authentications, program executions, network connections, and file manipulations) from endpoints and network appliances.
To discover potential intruders from these events, organizations predominantly rely on expert-written SIEM rulesets~\cite{alahmadi2022false,sommer2010outside,bhatt2014siem}.
Such rulesets implement the concept of \emph{misuse detection} (also called \emph{rule-based} or \emph{signature-based} detection), i.e., they contain definitions of known malicious behavior.
The main reasons to rely on misuse detection are its effectiveness in finding known attacks, simplicity of operation, and ability to ease investigation by providing alert details such as name and description of detected attacks~\cite{nist2007idps}.

However, these advantages come at the cost of potential detection blind spots, as rules typically fail to cover all possible variations of an attack.
Adversaries may thus \emph{evade} detection either by purposefully modifying attacks or inadvertently performing attacks in a subtly different way.
These detection blind spots are particularly dangerous when using publicly available rulesets, e.g., the widespread \emph{Sigma} rules~\cite{sigma}:
Due to public availability, adversaries can test attacks against rules and, in case of detection, modify them to evade any matches, thereby reducing the risk of being detected.
This issue of rule evasions is well known in related fields~\cite{ptacek1998insertion,wagner2002mimicry,cheng2012evasion,koret2015antivirus}, but has not been studied for SIEM rules until now.

In this paper, we show that widespread SIEM rules are indeed prone to evasions and present a novel method to detect such evasions with few or even no false alerts.
For this purpose, we first analyze 292 publicly available and widely used SIEM rules, showing that at least 44\,\% of them can be evaded using straightforward techniques, thus creating critical detection blind spots.
To remedy this situation, we introduce \emph{adaptive misuse detection}, which is an extension of (conventional) misuse detection that aims to detect SIEM rule evasions and otherwise undetected variants of known attacks in addition to (conventional) rule matches.
This is achieved by comparing incoming events to SIEM rules on the one hand and known-benign events on the other hand and deciding which of the two categories they are more similar to. %
Beyond that, if a potential evasion has been identified, adaptive misuse detection assesses which rule(s) were attempted to evade.
This \emph{rule attribution} aims to ease alert investigation for security  analysts by showing them potentially attempted attacks, similarly to conventional rule matches.

As a proof-of-concept implementation of adaptive misuse detection, we present \name{}, an open-source \textit{\textbf{A}}daptive \textit{\textbf{Mi}}suse \textit{\textbf{De}}tection \textit{\textbf{S}}ystem, which detects activity that is similar to known attacks, yet does not trigger the respective SIEM rules (i.e., evasions).
\name{} is based on supervised learning from SIEM rules versus benign events, allowing it to classify new events depending on their similarity either to known-malicious or known-benign activity.
Notably, by learning from existing SIEM rules, we overcome a common issue of supervised learning, namely, the necessity to manually create a comprehensive set of attacks for the training process.
For each potential evasion, \name{} suggests a list of SIEM rules that an adversary likely attempted to evade (rule attribution).

We evaluate \name{} using four weeks of SIEM events from an enterprise network with more than 50\,000 users and 512 manually generated evasions.
Our evaluation focuses on the most prevalent rule type in our SIEM ruleset (i.e., rules targeting malicious Windows process creations), but we also show that \name{} is applicable to other rule types as well.

At its default sensitivity, \name{} successfully detects a majority of our crafted evasions (70\,\%) without any false alerts, despite a realistically unbalanced validation dataset with \textasciitilde 145\,000 times as many benign as malicious events.
We also show that our approach of learning from SIEM rules is a better choice for practical application as compared to a conventional supervised approach that learns from attack events instead.
Finally, we show that \name{} is fast enough for operation in large enterprise networks and still performs reasonably in case of training data tainted with evasions.

In summary, to overcome the inherent problem of detection blind spots of SIEM rules while preserving the wished-for advantages of misuse detection, our contributions are:
\begin{itemize}[leftmargin=*]
\item We perform an analysis of almost 300 widespread open-source SIEM rules and show that approximately half of them can be evaded with minor effort, thus enabling adversaries to potentially avoid detection (Section~\ref{sec:analysis}).
\item We propose the concept of adaptive misuse detection, an extension of misuse detection that aims to detect SIEM rule evasions and undetected variants of known attacks by classifying events based on their similarity to either SIEM rules or historical benign events (Section~\ref{sec:amd}).
\item We design and implement \name{}, an open-source, proof-of-concept adaptive misuse detection system for practical application in enterprise networks (Section~\ref{sec:amides}).
\item We evaluate \name{} using data from a large enterprise network, showing its ability to detect a majority of our crafted SIEM rule evasions with zero false alerts and a computational performance that allows for real-world operation even in very large networks (Section~\ref{sec:evaluation}).
\end{itemize}

\section{Background: Detection of Cyberattacks in Enterprise Networks}
\label{sec:background}

Cybercriminals, state-sponsored groups, and other cyberattackers are causing vast damage by infiltrating enterprise networks and subsequently conducting data theft, espionage, and sabotage~\cite{verizon2022dbir,bitkom2021attacks}.
Among the victims are companies, authorities, and military institutions from all over the world~\cite{wikipedia2022incidents}.
Experience shows that purely preventive measures (such as timely patching, restrictive access control, and raised user awareness) are not sufficient to thwart all attacks~\cite{lord2022investments}.
Sooner or later, attackers usually succeed in finding a vulnerability they can exploit (be it technical or social), allowing them to gain a foothold in the network for subsequent operations~\cite{diogenes2018cybersecurity}.
Consequently, organizations need to establish comprehensive capabilities for detecting adversarial activity in their networks~\cite{wolsing2022ipal}.
These capabilities are crucial as detection is a prerequisite for reaction, i.e., stopping an attack, analyzing its impact, and attributing it to an adversary~\cite{vielberth2020soc}.

However, detecting adversarial activity in enterprise networks is a challenging task due to the sheer amount and noisiness of potential attack indicators~\cite{alahmadi2022false}.
Traditional sources of such indicators include alerts from firewalls, network-based intrusion detection systems, and anti-virus software~\cite{bhatt2014siem}.
Yet, since adversaries and malware increasingly make use of encryption and obfuscation techniques to evade traditional detection, more and more behavioral data need to be collected from endpoints, i.e., client and server computers~\cite{bridges2018host}.
Such behavioral data comprise low-level events such as process creations, network connections, and file manipulations, all of which are valuable indicators of malicious activity despite their large volume~\cite{acsc2021windows}.
To make sense of this vast amount of data, organizations utilize \emph{Security Information and Event Management} (SIEM) systems, which centrally collect, store, and analyze all security-related data~\cite{bhatt2014siem}.
SIEM systems usually collect data from log files or logging daemons/agents on the source systems, mostly in Syslog and Windows Event Log format.
Since their volume renders manual inspection impossible, SIEM systems need to automatically analyze the data for threats~\cite{alahmadi2022false}.
In case a potential threat is found, an alert is generated and needs to be sifted by human analysts, usually in the context of a security operations center~\cite{vielberth2020soc}.

The predominant method used by operational SIEM systems to automatically detect malicious activity is \emph{misuse detection} (also called \emph{rule-based} or \emph{signature-based} detection)~\cite{alahmadi2022false,sommer2010outside,bhatt2014siem}.
Misuse detection applies a set of \emph{rules} written by security experts to each collected event (or correlations thereof).
Each rule contains one or more \emph{signatures}, which describe conditions that trigger the rule (e.g., a regular expression matching a certain event field).
In addition, each rule usually has an expressive title, a description, and optionally references to further information on the respective attack.
Some SIEM systems also utilize \emph{anomaly detection} in addition to misuse detection~\cite{bhatt2014siem}.
However, experience shows that this method often produces prohibitive numbers of false alerts (cf. Section~\ref{sec:amides}).
Hence, misuse detection is currently the prime means for detecting cyberattacks in enterprise networks~\cite{alahmadi2022false}.

Still, misuse detection is not a silver bullet.
Research in the related fields of network-based and host-based intrusion detection as well as malware detection has shown that rules can often be \emph{evaded}, i.e., attacks can be performed successfully without triggering a rule by slightly modifying the attack, e.g., by inserting dummy characters into malicious strings to avoid matching a signature~\cite{ptacek1998insertion,wagner2002mimicry,cheng2012evasion,koret2015antivirus}.
Common reasons for evadable rules are that (1)~the rule author tailored the rule to a concrete attack and did not anticipate all possible variations and obfuscations and (2)~matching all possible variations and obfuscations would make the rule (or the underlying analysis engine) too complex or computationally infeasible.
Therefore, evasions are a fundamental and inherent problem of misuse detection, which leads to \emph{detection blind spots}, i.e., attacks may remain undetected despite appropriate rules being deployed~\cite{cheng2012evasion}.
Note that an evasion can either be purposeful (i.e, an attacker anticipates that a certain rule is deployed and deliberately evades it) or incidentally (i.e., an attacker unknowingly performs an undetected \emph{attack variant}, e.g., by executing a command line with slightly different arguments than covered by the respective rule).
For brevity, in the remainder of this paper, we subsume both kinds under the term \emph{evasion}, thereby meaning a purposeful or incidental evasion of a SIEM rule, including undetected variants of known attacks.

To the best of our knowledge, there is no research yet that analyzes SIEM rules for potential evasions.
Possibly this is due to the fact that there has been no large public corpus of SIEM rules for a long time.
However, this changed significantly with the advent of \emph{Sigma}, which is a generic and open signature format for SIEM systems~\cite{sigma}.
In the last few years, following the general trend of open-source information security~\cite{lambert2019githubification}, Sigma has become a highly active project with a large number of contributors~\cite{sigma,opensourcesecurityindex}.
While this trend enables organizations to make use of a large corpus of comprehensive and up-to-date SIEM rules, it also bares the risk of purposeful evasions in case an adversary correctly assumes that public Sigma rules are being used by a potential victim.

In the following, we will show that Sigma rules are indeed prone to evasions and thus measures for evasion detection are required to reduce critical detection blind spots in enterprises.

\section{Analysis of SIEM Rules for Evasions}
\label{sec:analysis}

\begin{table*}
  \caption{Almost half of the analyzed SIEM rules (129 of 292) can be evaded using the five straightforward evasion types presented in this table (each with one concrete example), thus causing critical detection blind spots in enterprise networks.}
  \small
  \centering
  \begin{tabularx}{\textwidth}{llllX}
    \toprule
    \textbf{Evasion type} & \textbf{Sample affected rule} & \textbf{Affected search term} & \textbf{Sample match} & \textbf{Sample evasion} \\
    \midrule
    Insertion & win\_susp\_schtask\_creation & * /create * & schtasks.exe /create ... & schtasks.exe /"create" ... \\
    \cellcolor{Gray}Substitution & \cellcolor{Gray}win\_susp\_curl\_download & \cellcolor{Gray}\textvisiblespace-O\textvisiblespace & \cellcolor{Gray}curl -O http://... & \cellcolor{Gray}curl -{}-remote-name http://... \\
    Omission & win\_mal\_adwind & *cscript.exe *Retrive*.vbs * & cscript.exe ...\textbackslash{}Retrive.vbs & cscript ...\textbackslash{}Retrive.vbs \\
    \cellcolor{Gray}Reordering & \cellcolor{Gray}win\_susp\_procdump & \cellcolor{Gray}* -ma ls* & \cellcolor{Gray}procdump -ma ls & \cellcolor{Gray}procdump ls -ma \\
    Recoding & win\_vul\_java\_remote\_dbg & *address=127.0.0.1* & ...address=127.0.0.1,... & ...address=2130706433,... \\
    \bottomrule
  \end{tabularx}
  \label{tab:evasions}
\end{table*}

To lay the foundation for our work, we analyze a representative set of SIEM rules with respect to potential evasions.
For this purpose, we chose a subset of \emph{Sigma rules}, which are probably the most widely used corpus of open-source SIEM rules at the moment.
In the following, we give a short introduction to Sigma, describe our analysis goals and methodology, and finally present our findings, showing that evasions indeed induce significant detection blind spots.

\paragraph{Introduction to Sigma Rules}

Sigma is a generic and open signature format for SIEM systems.
It allows for flexible rules in YAML format that can detect malicious or suspicious behavior in any type of text-based log data.
Sigma rules can be automatically converted to queries for common SIEM products (e.g., Splunk or Elasticsearch).
Aside from these conversion tools and Sigma's specification, its GitHub repository~\cite{sigma} contains a large corpus of detection rules, which are continuously revised and extended by a large community.
According to the Open Source Security Index, the Sigma project is among the most popular and fastest growing open source security projects on GitHub (sixth place overall, highest ranked project with detection focus as of January 2023)~\cite{opensourcesecurityindex}.
In our professional experience, Sigma rules are widely used by organizations in practice.

\paragraph{Analysis Goal and Methodology}

The goal of our analysis was to quantify the risk of detection blind spots for existing Sigma rules by finding concrete evasions for them.
We had to restrict our analysis to a subset of rules from the Sigma repository due to the high effort of manual analysis and the fact that some types of rules act on log data of software or hardware that is not generally available (e.g., commercial cloud services or appliances).
We thus chose to analyze the subset of rules that act on \emph{process creation events} on Windows systems.
This rule type is the most frequent (making up 41\,\% of all rules at the time of analysis) and does not depend on log data from proprietary products except for Windows.
Furthermore, process creation events are known to be a valuable source for threat detection~\cite{acsc2021windows}.
They comprise information on newly created processes and are generated either by Windows itself~\cite{ultimateid4688} or by the tool Sysmon~\cite{russinovich2022sysmon,ultimateid1}.
The contained fields most often searched by the Sigma process creation rules are \texttt{CommandLine} (the full command line of the process creation; 45\,\% of all search expressions in the considered rules), \texttt{Image} (the full path of the executed image; 29\,\% of search expressions), and \texttt{ParentImage} (the full path of the parent process image; 10\,\% of search expressions).
There are \textasciitilde 20 more fields that are searched by only a few rules, e.g., \texttt{Description}, \texttt{ParentCommandLine}, and \texttt{User}.

The analysis was conducted as follows.
We analyzed all process creation rules that were contained in the Sigma repository on February 4, 2021 (commit ID 12054544).
First, we reviewed each rule in detail, including potential references given within the rule (e.g., threat reports describing the malicious behavior that should be detected by the rule).
Next, we tried to re-enact the malicious process creation as described by the rule on a Windows~10 system (e.g., by running \texttt{powershell.exe~/C Clear-EventLog System}).
We manually reviewed the Windows event log to verify our commands and then ran scripts to check for Sigma rules matching these events.
In case we succeeded to match the current rule, we then tried to find command lines that perform the exact same action, but without matching the rule (i.e., evasions).

For this to work, we assumed that an attacker who can create a process with command line arguments is also able to alter these arguments (e.g., \texttt{curl.exe -{}-remote-name example.com} instead of \texttt{curl.exe -O example.com}).
We made sure that successful evasions did not match any additional Sigma rules (i.e., rules not triggered by the original match).
Finally, we assigned one of four labels to each rule: (1)~\emph{full} if we were able to completely evade the rule, (2)~\emph{partial} if the rule contains OR-branches of which we could evade at least one, but not all, (3)~\emph{none} if we could not find an evasion, and (4)~\emph{broken} if we found the rule to be faulty.

\paragraph{Analysis Results}

Of the 292 analyzed Sigma rules, we found that 110 (38\,\%) can be \emph{fully} evaded and 19 (7\,\%) can be \emph{partially} evaded.
For another 51 rules (17\,\%), we found that evasion might be possible but could not confirm any concrete evasion instances either due to unavailable target software (mostly malware) or excessive effort for conclusive analysis.

We achieved all evasions by adapting the processes creation command line in multiple ways.
For this purpose, we found a total of five evasion types during our analysis as exemplified in Table~\ref{tab:evasions}: (1)~\emph{Insertion} of ignored characters into the command line (e.g., double quotes or spaces), (2)~\emph{substitution} of synonymous characters or arguments (e.g., a hyphen instead of a slash before an argument), (3)~\emph{omission} of unnecessary characters (e.g., shortening arguments), (4)~\emph{reordering} of arguments, and (5)~\emph{recoding} of arguments.
We created at least three matching and three evading events for each evadable rule to achieve variability, ending up with 461 matches and 512 evasions in total, which were later used in the context of our evaluation (cf. Section~\ref{sec:evaluation}).
Summarizing, we were able to evade almost half of the rules, each with multiple variants.

We would like to emphasize that generally our results come as no real surprise and are not necessarily specific to the analyzed Sigma rules.
Instead, evasions are an intrinsic problem of misuse detection due to the impracticality of covering every possible mutation with hard-coded signatures (cf. Section~\ref{sec:background}).
This makes it practically impossible to ``fix'' the affected rules in the sense of adapting them to detect all possible evasions.
However, a subsidiary result of our analysis is that we found 12 of the 292 rules to be \emph{broken}, i.e., failing to detect what was intended by the rule author.
We excluded these rules from our evasion analysis.
Furthermore, we provided fixes to the Sigma maintainers for all broken rules that had not yet been fixed or removed by the Sigma community in the meantime, resulting in four rules that were fixed through our feedback.

In conclusion, we find that the risk of detection blind spots through rule evasions is indeed high for the analyzed Sigma rules, which are widely used in practice.
Even small attack mutations using simple techniques suffice to evade detection.
Consequently, adversaries might remain undetected despite performing commonly known attacks.
In the following, we therefore present our main idea, \emph{adaptive misuse detection}, which aims to solve this dilemma of easily-evadable rules by classifying incoming SIEM events based on their similarity to SIEM rules versus historical benign events.

\section{The Case for Adaptive Misuse Detection}
\label{sec:amd}

\begin{figure}[h]
  \centering
  \includegraphics[width=\linewidth]{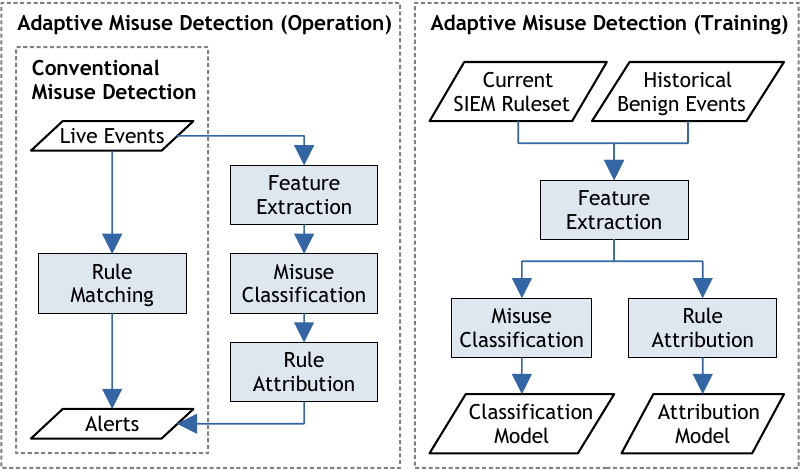}
  \caption{Our concept of adaptive misuse detection extends conventional misuse detection by components that aim to detect rule evasions and attack variants. For this purpose, classifiers are trained on SIEM rules versus benign events.}
  \label{fig:concept}
\end{figure}

The results of our analysis lead to the question of how the discovered detection blind spots can be remedied.
Therefore, we propose \emph{adaptive misuse detection}, a novel methodology to significantly reduce blind spots by detecting rule evasions \emph{in addition} to conventional rule matches.
The components of our methodology are depicted in Figure~\ref{fig:concept} and described in the remainder of this section, followed by a conceptual comparison with related detection approaches.

\paragraph{Rule Matching}
First of all, adaptive misuse detection is an extension of (conventional) misuse detection with additional machine learning components (described below).
In other words, conventional rule matching is still performed and matching rules still trigger alerts.
This is because the machine learning components are only applied to a subset of incoming events where evasion detection is sensible (cf. Section~\ref{sec:amides}).
Hence, adaptive misuse detection executes rule matching and the machine learning components in parallel and then merges the resulting alerts.
This design ensures that no attack is missed that conventional misuse detection would discover, yet in addition evasions can be detected (which are, by definition, missed by conventional misuse detection).

\paragraph{Rationale for Evasion Detection}
Adaptive misuse detection is based on the assumption that SIEM events of successful evasions are still very similar to those of the original attack.
In our experience, an analyst can often quickly spot an evasion when reviewing the respective event.
This is because the SIEM events that are predominantly used for attack detection (i.e., endpoint events, cf. Section~\ref{sec:background}) are captured at kernel level,
immediately before or after the respective action is executed by the operating system.
At this level, most obfuscations (as often seen in malware binaries or scripts~\cite{aslan2020malware,bohannon2019dosfuscation,bohannon2017revoke}) must already be resolved, else the operating system would not be able to execute the desired action.
For example, in the case of a malicious process execution, the executable name (e.g., \texttt{powershell}) ultimately has to be passed to the operating system in plaintext for execution.
The same holds for command line parameters of executed processes (e.g., \texttt{set-executionpolicy unrestricted}), which ultimately need to be parsed by the started process, thereby restricting the set of possible evasions in the captured SIEM events.

As a consequence, a promising approach for evasion detection is to look for events that are \emph{similar} to one or more signatures contained in the ruleset.
However, there is an important constraint, which is the \emph{avoidance of false alerts}:
Large enterprise networks often deal with millions or billions of SIEM events per day~\cite{bhatt2014siem} and typically, only a small number of analysts handle security alerts generated by multiple security systems~\cite{sans2019soc}.
Many analysts report to be overwhelmed by false alerts, impairing their ability to discover actual attacks~\cite{alahmadi2022false,sans2019soc}.
Since the number of benign events is usually several orders of magnitude larger than the number of attack-related events~\cite{axelsson2000base}, a practical approach thus requires an extremely small false positive rate.
Consequently, when aiming to detect evasions based on their similarity to rules, it is crucial to avoid an overly broad definition of similarity.

\paragraph{Misuse Classification}
To unite these conflicting requirements of detecting as many evasions as possible while at the same time keeping false alerts low, we propose a machine learning based approach for finding potential evasions, i.e., events that are similar to SIEM rules but do not match them.
More precisely, we propose to apply supervised learning to classify whether a new event is more similar to \emph{deployed SIEM rules} or to \emph{historical benign events}.
In other words, we propose to train classifiers with features extracted from SIEM rules versus features extracted from benign events, both taken from the network where the system will be operated.
We call this step \emph{misuse classification}.
To the best of our knowledge, this approach has not been studied before (cf. Section~\ref{sec:relatedwork}).

Notably, the idea of learning from already-existing rules overcomes a common issue of supervised learning, namely, the necessity to create a comprehensive set of attacks for the training process.
Instead, the knowledge of what is malicious is taken from the ruleset, which -- in case of Sigma -- is publicly available, comprehensive, and regularly updated by a large community.
We decided to call our approach \emph{adaptive} misuse detection since it adapts to the target environment by training against its benign activity, thereby adjusting its features and sensitivity to properly distinguish attacks from benign events.
Moreover, the approach allows to adjust its sensitivity (better detection rate or less false alerts), which is important in practice because the number of acceptable (false) alerts depends strongly on the number and workload of available analysts.
Overall, misuse classification aims to enable a pinpoint detection of evasions while minimizing false alerts.

\begin{table*}
  \caption{Enhancements of adaptive misuse detection over conventional misuse detection}
  \small
  \centering
  \begin{tabularx}{\textwidth}{ p{2.0cm} p{6.5cm} X } %
    \toprule
    & \textbf{Misuse Detection} & \textbf{Adaptive Misuse Detection} \\
    \midrule
    \textbf{Goal}
    & Detect known malicious behavior
    & Additionally detect rule evasions and attack variants \\
    \multirow{2}{*}{\cellcolor{Gray}\textbf{Rationale}}
    & \cellcolor{Gray}Malicious behavior is explicitly defined, everything else is considered benign.
    & \cellcolor{Gray}Behavior is additionally classified by its similarity to known malicious versus known benign behavior. \\
    \textbf{Method}
    & Matching of expert-written rules
    & Additional classifiers trained on rules versus benign events \\
    \cellcolor{Gray}\textbf{Customization}
    & \cellcolor{Gray}Addition, removal, or customization of rules
    & \cellcolor{Gray}Additional adjustment of sensitivity \\
    \textbf{Alert details}
    & Matching rule(s) name, ID, and further attack details
    & Identical for rule matches; estimate for potentially evaded rules \\
    \bottomrule
  \end{tabularx}
  \label{tab:concept}
\end{table*}

\paragraph{Rule Attribution}

A substantial reason for the popularity of (conventional) misuse detection is that each rule contains an expressive title, description, and possibly references to further information, which greatly eases alert investigation.
Technically, we can say that each alert is \emph{attributed} to one or more rules.
This advantage is lacked by many machine learning based systems, which often generate alerts containing no further information on the potential attack~\cite{sommer2010outside}.

Since adaptive misuse detection learns from SIEM rules (versus benign events), information on which features represent which rule is available during training.
Consequently, when a potential evasion is detected by the misuse classification step, adaptive misuse detection allows to perform a further step that we call \emph{rule attribution}:
Even though an evasion, by definition, does not match any rule, we can still compare a potential evasion to all SIEM rules and thus estimate which rule(s) an adversary \emph{probably} evaded.
For example, when a command line is executed that is very ``similar'' to a signature of some rule (e.g., differing only in one parameter -- see Section~\ref{sec:amides} for details), the respective rule should be proposed to the analyst within the alert as potentially being evaded.

Note that depending on the concrete evasion, there can be several rule candidates due to overlapping or redundant contents.
Therefore, the rule attribution does not just output the most likely evaded rule but a \emph{ranked list} of potentially evaded rules, which is then added to the alert to ease investigation.
Even though the rule attribution is inevitably uncertain, it may still significantly help analysts to verify if an alert is true and better understand the context of an attack.

The idea of adaptive misuse detection and its differences to (conventional) misuse detection are summarized in Table~\ref{tab:concept}.
The training process will be further described in Section~\ref{sec:amides}.

\paragraph{Conceptual Comparison to Related Detection Approaches}
Adaptive misuse detection has some properties that make it very promising for real-world application.
Compared to \emph{conventional misuse detection}, its obvious benefit is the ability to detect evasions.
This benefit comes at the expense of a higher operational and computational effort:
Benign events must be extracted from the target network first to then perform the training.
This training should be repeated when SIEM rules change or when the benign behavior changes significantly.
However, as most organizations already utilize conventional misuse detection, adaptive misuse detection constitutes a straightforward way to reduce detection blind spots.

Adaptive misuse detection also shares conceptual properties with \emph{anomaly detection}, in the sense that historic events are used for training the system.
However, while anomaly detection methods learn from unlabeled, historic events \emph{only} to then detect everything that is \emph{not normal}~\cite{nist2007idps}, adaptive misuse detection leverages comprehensive and ever-growing rulesets (such as Sigma) of known-malicious behavior.
Anomaly detection systems, on the other hand, are built on the assumption that attacks correspond to anomalous behavior in the trained model and benign activity corresponds to normal behavior~\cite{gates2006challenging}.
This assumption has repeatedly been questioned and criticized~\cite{gates2006challenging,sommer2010outside}:
Since anomaly detection systems have no knowledge about what is malicious, they tend to produce an unacceptable number of false alerts in real-world applications, thus causing threat alert fatigue that can lead to missed attacks~\cite{alahmadi2022false,hassan2019nodoze,sans2019soc}.
Even worse, they can also miss actual attacks because they are not necessarily being recognized as anomalous~\cite{gates2006challenging}.
Furthermore, (adaptive) misuse detection has the significant advantage of providing alert details such as the relevant rule's name and description, whereas anomaly detection only reports that \emph{something} unusual has happened~\cite{nist2007idps}.
This makes root cause analysis much harder, especially in large networks with many users and systems.

Another approach to detect malicious events is \emph{conventional supervised learning} in the sense that a classifier is trained on malicious versus benign events.
The problem with this approach is its dependence on a comprehensive, realistic, and up-to-date set of malicious events (i.e., logged attacks).
Experience shows that such data are extremely difficult and labor-intense to create and maintain~\cite{uetz2021socbed}, thus casting doubt on the practical utility of this approach~\cite{sommer2010outside}.
Adaptive misuse detection avoids this issue by utilizing community-maintained SIEM rules as malicious samples, which are inherently comprehensive, realistic, and up-to-date.

Finally, there is the idea of \emph{synthetic signature generation}, i.e., similar signatures are automatically created from existing ones to detect variants of known attacks~\cite{sohi2021rnnids}.
The systematic problem with this approach is that it may result in prohibitive numbers of false alerts since there is no tuning process to automatically adjust sensitivity to a concrete environment.
Adaptive misuse detection avoids this problem by training against benign data from the concrete environment to automatically find sensible features and decision thresholds.

In conclusion, our discussion indicates that adaptive misuse detection is a promising approach for detecting SIEM rule evasions in enterprise networks, thus reducing critical detection blind spots.
Yet, it remains to show that its theoretical benefits can be transferred to practical application.

\section{\name{}: Effective and Efficient Detection of SIEM Rule Evasions}
\label{sec:amides}

\begin{figure*}
  \centering
  \includegraphics[width=\textwidth]{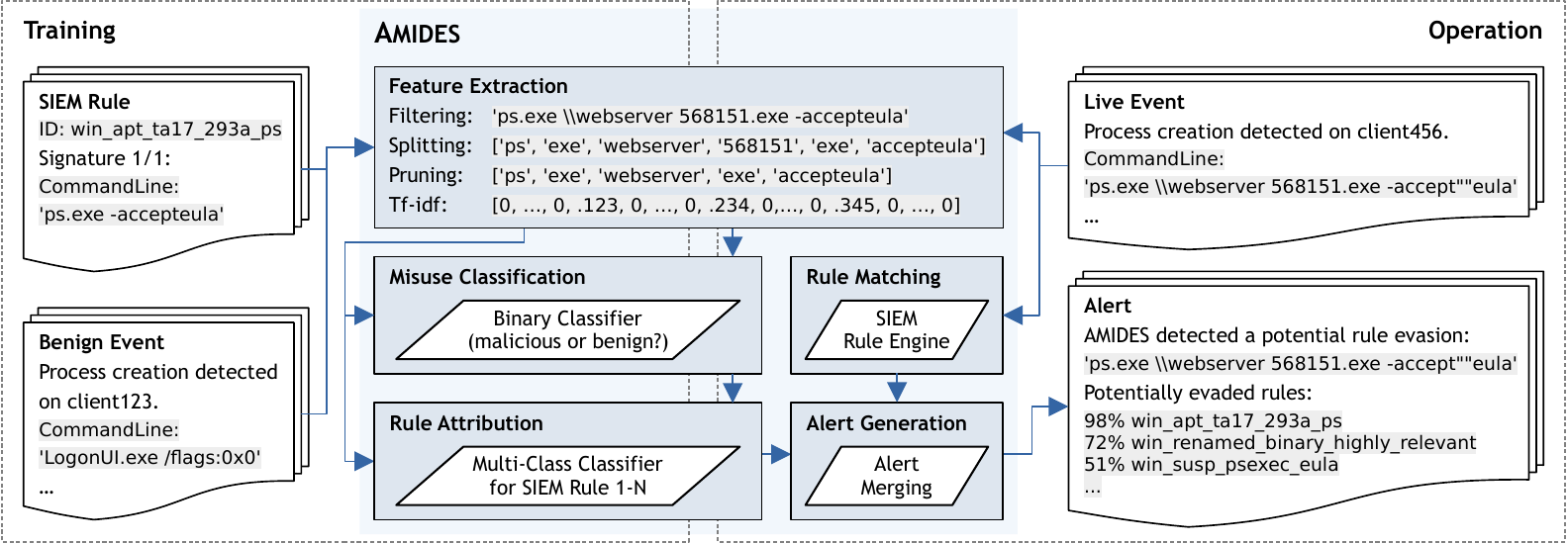}
  \caption{\name{} is a proof-of-concept adaptive misuse detection system aiming to detect attacks that evade SIEM rules as well as otherwise undetected attack variants in addition to conventional SIEM rule matches. It learns from SIEM rules on the one hand and historical benign events on the other hand and can thus even estimate which SIEM rule an adversary attempted to evade.}
  \label{fig:amides}
\end{figure*}

To verify the practical applicability and benefits of adaptive misuse detection, we designed and implemented \name{}, a proof-of-concept \textit{\textbf{A}}daptive \textit{\textbf{Mi}}suse \textit{\textbf{De}}tection \textit{\textbf{S}}ystem.
Our main goals are (1) the detection of as many evasions as possible (effectiveness) and (2) false positive rates and resource requirements that are both low enough for real-world operation of the system in large enterprise networks (efficiency).
In the following, we give a brief overview of the system and then describe its principal components in detail, finishing with a consideration of the computational performance.

\paragraph{Overview}

\name{} is a proof-of-concept implementation of adaptive misuse detection, as proposed in Section~\ref{sec:amd}.
It is meant to extend conventional misuse detection components of SIEM systems in enterprise networks.
As described, this implies that conventional rule matching will still be performed while additional machine learning components aim to detect rule evasions.
Figure~\ref{fig:amides} shows an overview of \name{}, including exemplary input data for both training and operation.
\name{} is trained using the current SIEM ruleset and historical benign events of an organization.
The training should be repeated when either the SIEM ruleset or the benign activity change significantly (e.g., due to major software changes).

During operation, incoming SIEM events are passed to the rule matching component (as for conventional misuse detection) \emph{and} the feature extraction component, which generates a feature vector from the event that is suited for the subsequent components.
This vector is first passed to the misuse classification component, which classifies the event as either malicious or benign.
If classified as malicious, the feature vector is also passed to the rule attribution component, which generates a ranked list of rules potentially evaded by the event.
Finally, potential alerts of the rule matching and machine learning components are merged by the alert generation component to create one single alert per malicious event.

Since we limited our detailed Sigma analysis (cf. Section~\ref{sec:analysis}) to Windows process creation rules due to the high manual effort of creating evasions, we also focused on this rule/event type during the design of \name{} (e.g., influencing its feature engineering).
However, \name{} is also applicable to other types:
While our evaluation focuses on process creations (cf. Section~\ref{sec:evaluation}), we also prototypically evaluated three additional Sigma rule types that we found to be inherently evadable, namely, those addressing web requests, Windows registry changes, and PowerShell scripting (cf. Section~\ref{sec:rule-types}).

\name{} is designed and implemented with performance in mind for application in large enterprise networks (cf. Section~\ref{sec:speed-and-robustness}).
We implemented \name{} in Python, with principal third-party packages being \texttt{sklearn} and \texttt{numpy}.
\name{} is freely available under the GPLv3 license~\cite{amides}.
In the following, we describe the system in more detail.

\paragraph{Input Data and Feature Extraction}

According to our concept of adaptive misuse detection, \name{} requires a SIEM ruleset and a set of benign events for the training process.
Since we assume that organizations adopting \name{} are already using conventional misuse detection (cf. Section~\ref{sec:background}), they can simply use their current SIEM ruleset with \name{}.
For now, \name{} only supports Sigma rules, but other formats such as Splunk could be added in the future.
In addition, \name{} also requires a set of benign events for the training process.
Since organizations store large numbers of events in their SIEM systems anyway (usually for a few days up to several years), the required events can be obtained by extracting the latest few days or weeks from the SIEM system.
Ideally, these events should not contain any attacks.
However, in practice, the data might be tainted with undetected attacks (including evasions), which we consider in our evaluation (cf.~Section~\ref{sec:speed-and-robustness}).
We thus recommend to sift the benign dataset before training and only update it when required.

Features are extracted from SIEM events using established methods to obtain an appropriate input vector for the machine learning components (cf. Figure~\ref{fig:amides}):
First, relevant fields (here: \texttt{CommandLine}) are extracted and specific characters are removed that can be used to evade signatures (regex \texttt{["'\textasciicircum{}`´]}).
The resulting string is split by non-word characters (regex~\texttt{\textbackslash W}), yielding a sequence of terms.
Then, a pruning step is applied to remove likely irrelevant terms such as timestamps and unique IDs (more precisely: decimal and hexadecimal numbers over three digits as well as strings over 30 characters).
Finally, the term sequence is transformed into a weighted numeric vector using tf--idf~\cite{tfidf}, thereby assigning higher values to rarely occurring terms.
The calculated idf values are stored for later use during the operation phase, in which the exact same feature extraction is applied to live events.
For further discussion on our implementation choices see Appendix~\ref{sec:experiments-description}.

The feature extraction from Sigma rules is more complex.
Each rule contains at least one signature, consisting of one or more \emph{search identifiers}, coupled by logical conditions, where each search identifier can contain required keys, values, and modifiers such as \texttt{endswith}.
An exemplary signature is \texttt{Image|endswith: "\textbackslash{}schtasks.exe" AND CommandLine: "*~/create~*" AND NOT User: "NT~AUTHORITY\textbackslash{}SYSTEM"}.
For now, \name{} performs a simplified signature parsing that extracts from a rule all strings being searched for in an event's command line, as long as they are not part of an exclusion (i.e., within a \texttt{NOT} branch of a signature).
While this implementation does not cover Sigma's full potential, it suffices to capture the essence of most process creation rules, as shown by our evaluation (cf.~Section~\ref{sec:evaluation}).
After extraction, the search strings are processed by the feature extraction component in the same way as the command lines from SIEM events (cf. Figure~\ref{fig:amides}).

\paragraph{Misuse Classification and Rule Attribution}

\name{} implements the two machine learning components proposed in our adaptive misuse detection concept, namely, misuse classification and rule attribution.
The former decides if an incoming event is malicious or benign whereas the latter enriches an evasion alert with a ranked list of SIEM rules that were likely evaded by the event (cf. Section~\ref{sec:amd}).
The misuse classification component consists of a supervised binary classifier, which is trained with all feature vectors extracted from the SIEM rules (labeled as malicious) versus all feature vectors extracted from the benign events (labeled as benign).
During training, \name{} calculates a sensible decision threshold range and maps it to $[0,1]$ to allow for intuitive configuration of its sensitivity even by non-expert users, with $0.5$ being the default value (see Appendix~\ref{sec:experiments-description} for more details).

The second machine learning component, i.e., rule attribution, consists of multiple supervised binary classifiers, more precisely, one per SIEM rule of the supplied ruleset.
Each classifier is trained with the feature vectors extracted from the respective rule (labeled as malicious) versus all benign events (labeled as benign).
During operation, for each incoming event classified as malicious by the misuse classification component, the rule attribution runs all its classifiers and creates a ranked list of classifier confidence values.
The more similar a rule is to the event, the higher is its confidence value and the higher it appears in the list.
By default, the top 10 rules are shown to allow analysts a comprehensive investigation without overwhelming them with likely irrelevant rules.

On a technical level, we utilize support vector machines (SVMs)~\cite{cortes1995svm} as classifiers for both misuse classification and rule attribution since they are well suited for our highly imbalanced datasets~\cite{tang2009svms} of high-dimensional, sparse feature vectors~\cite{joachims1998text}.
Training and hyperparameter tuning are performed using established methods, i.e., grid search with stratified $5$-fold cross validation, optimizing for $F_1$ score.
Our implementation choices are further discussed in Appendix~\ref{sec:experiments-description}.

\paragraph{Rule Matching and Alert Generation}

The rule matching component performs the task of conventional misuse detection, i.e., matching incoming events against the chosen SIEM ruleset.
As described in Section~\ref{sec:amd}, this component complements the machine learning components, since the latter focus on specific SIEM event types that are prone to evasions (e.g., process creation events).
Technically, we utilize the open-source log processor Logprep~\cite{logprep} for Sigma rule matching.

The subsequent alert generation component simply merges alerts of the rule matching component and the machine learning components into just one alert per event.
Consequently, an alert can contain none, one, or multiple rule matches as well as none or one evasion alert, in this case including the top 10 most likely evaded rules (cf. Figure~\ref{fig:amides}).

\paragraph{Performance Considerations}

A detection system is pointless if its performance requirements are infeasible under real-world conditions.
Hence, \name{} implements several measures to enable high event throughput.
In particular, we assume that the number of benign events in operation is orders of magnitude larger than the number of attack-related events and thus optimize for high benign event throughput.
In addition, since many users in an organization often perform the same tasks (e.g., opening a browser), the fraction of unique actions is usually quite small (cf. Section~\ref{sec:evaluation}).
Therefore, we implement an in-memory cache to avoid repeated classification of already-seen feature vectors.
The cache size is configurable and \name{} periodically outputs performance metrics to monitor cache utilization and efficiency (hits and misses).
All results of the misuse classification and rule attribution steps are added to the cache along with the associated feature vector, thus alerts can be reproduced from the cache instead of repeating the costlier classification steps.

Secondly, the pruning step of the feature extraction further reduces the number of unique command lines because it removes variable, yet irrelevant parameters such as unique IDs from otherwise frequent command lines, thus making caching more efficient.
Thirdly, since the rule attribution step requires to run one classifier \emph{per rule per event} whereas the misuse classification step only involves one classifier \emph{in total per event}, the rule attribution is only executed if an event is classified as malicious by the misuse classification step.
Lastly, we implemented performance optimizations for the training phase as well.
For example, \name{} deduplicates benign feature vectors before training, which greatly reduces SVM training time without impairing classification performance.
Furthermore, instead of performing an SVM parameter optimization for each rule attribution classifier, we inherit the optimized parameters from the misuse classifier, since its training set is a superset of the rule attribution training sets.

Without such optimizations, operation in large enterprise networks would likely be infeasible.
To show that \name{} can handle realistic event volumes, we present performance measurements as part of our evaluation in the following.

\section{Evaluation}
\label{sec:evaluation}

\name{} aims to significantly reduce detection blind spots in enterprise networks by automatically detecting SIEM rule evasions.
However, good detection performance alone is not sufficient to be beneficial under real-world conditions.
In this section, we measure key properties of \name{} to evaluate its overall practical value.
More precisely, we subdivide our evaluation into three research questions (RQ) of interest:

\textbf{RQ1:} \emph{How well does \name{} detect SIEM rule evasions?}
To address this question, we apply \name{} to a realistic dataset, quantify true and false classifications, and present a practical interpretation of the results.
Moreover, we compare it to a benchmark approach that is trained with attack events instead of SIEM rules against the same benign events.

\textbf{RQ2:} \emph{How accurate is the rule attribution?}
While RQ1 focuses on binary detection performance (i.e., whether an alert is raised or not), \name{} additionally suggests SIEM rules that might have been evaded by an adversary.
We evaluate the accuracy of this rule attribution by comparing its output against ground truth data, i.e., we check if the truly evaded rule is suggested and how high it is ranked within the suggestions.

\textbf{RQ3:} \emph{Is \name{} suited for real-world operation?}
Practical operation induces requirements beyond good classification and attribution performance.
We analyze three important aspects:
First, we assess if \name{}' training and operation are fast enough for typical enterprise event volumes.
Secondly, we check if \name{} still performs reasonably under tainted training data that inadvertently contain evasions.
Thirdly, we evaluate \name{}' generalizability to different event types.

Answering the three research questions requires realistic data to evaluate against, which we describe in the following.

\paragraph{Datasets and Ethical Considerations}

To the best of our knowledge, there exists no dataset yet that contains SIEM events with rule evasions or undetected attack variants.
We thus decided to create an own dataset.
For this purpose, we had the opportunity to collect SIEM events from a large enterprise network with more than 50\,000 users.
The data comprise four weeks of process creation events from May 2022, collected from Windows machines running a remote-controlled browser system and associated applications (such as PDF and image viewers) for all users of the enterprise.
In total, we collected 155\,373\,027 events, containing 266 unique executable filenames.
We processed the data only within the premises of the enterprise and only exported evaluation results.
All user names were already pseudonymized.
We carefully sifted the events to make sure they do not contain any attacks.
Thus, we use them as \emph{benign} events in the course of our evaluation.
For further details on the dataset collection and analysis process, see the supplementary document in our code repository~\cite{amides}.

Since the dataset cannot be shared with the research community due to security and privacy concerns, we created an additional set of benign events using the open-source testbed SOCBED~\cite{uetz2021socbed}, thus enabling researchers and potential users to test \name{}.
However, in this paper we solely present results for the enterprise dataset since the benign events of the testbed dataset are synthetically generated by a simple user emulation and therefore not as realistic.
Results for the synthetic dataset are available with the \name{} code to facilitate reproducibility~\cite{amides}.
Our set of evasions will not be made freely available due to ethical concerns, but can be shared with other researchers upon personal request.

Next, we describe how we utilize the enterprise dataset and our evasions for training and validating \name{}.

\paragraph{Training and Validation}

For the purpose of our evaluation, we first trained \name{} with a set of benign events versus a set of SIEM rules as described in Section~\ref{sec:amides}.
Then, to simulate live operation and validate \name{}' detection performance, we provided \name{} with a disjoint set of benign events versus a set of events containing evasions.
We used Weeks 1 and 2 of the enterprise dataset for training and Weeks 3 and 4 for validation, which is more realistic than a randomized partitioning since in real-world operation, the training is always performed on historic data and detection is then performed on future events.
As SIEM rules (training) and evasion events (validation), we used the Sigma process creation rules and the evasion events that we crafted for these rules as described in Section~\ref{sec:analysis}, respectively.
Note that the resulting training and validation datasets are highly imbalanced on purpose to reflect realistic ratios of benign activity to attacks~\cite{alahmadi2022false}.

Since our validation dataset is semi-synthetic (because the evasion events were created in a testbed environment), we need to make sure that \name{} cannot inadvertently learn from artifacts of the dataset generation process.
(Note that we could not generate the evasions within the enterprise network due to security concerns.)
Therefore, in the course of our evaluation we decided to solely use the \texttt{CommandLine} field as input to \name{}' feature extraction, since other fields could indicate whether an event stems from the enterprise network (and is thus benign).
For example, computer names differ between benign and evasion events, so the model could utilize them for classification, which would be a form of unwanted \emph{data snooping}~\cite{arp2022dos}.
Since all of our evasions are based on command line manipulations (cf. Section~\ref{sec:analysis}), restricting the input to this field still enables \name{} to detect all evasions.
In practice, more fields could be used if deemed useful for further improving the detection performance.

In the following, we present the results of our measurements concerning \name{}' training and validation process.

\subsection{\name{} Detects Most SIEM Rule Evasions without Any False Alerts (RQ1)}
\label{sec:classification-performance}

To answer the first research question, we now present and discuss \name{}' classification results.
Afterwards, we compare these results to those of a benchmark approach.

\paragraph{Classification Performance}

In the following, we present results for \name{}' misuse classification step.
Since its sensitivity is configurable, we start with results for the default threshold and then discuss the influence of threshold changes.

In absolute numbers, \name{} detected 358 events containing evasions (true positives) and missed 154 (false negatives), so 70\,\% of our evasions were successfully detected.
\name{} correctly ignored 74\,431\,740 benign events (true negatives) and raised zero false alerts (false positives).
Hence, \name{} clearly fulfills our classification goals (cf. Section~\ref{sec:amides}), namely, detecting many evasions (that would otherwise remain undetected) without overwhelming analysts with false alerts.
It thus achieves a significant reduction of detection blind spots as revealed by our SIEM rule analysis.
Notably, the false positive rate (i.e., fraction of incorrectly detected benign events) is zero despite the realistically unbalanced validation set with \textasciitilde 145\,000 times as many benign as malicious events.

\begin{figure}
  \includegraphics[width=\linewidth]{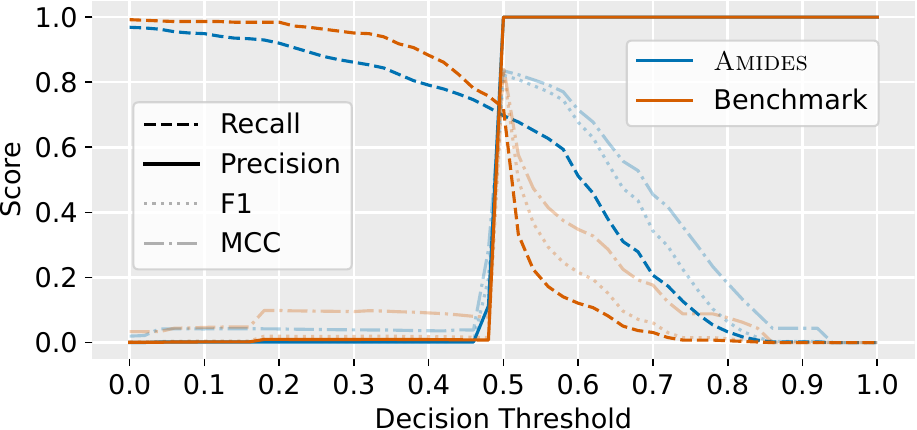}
  \caption{\name{} detects most SIEM rule evasions (70\,\%) with zero false alerts at default sensitivity ($0.5$), thus significantly reducing detection blind spots in enterprise networks. It keeps up with an (impractical) benchmark trained on attack events, showing the eligibility of learning from SIEM rules.}
  \label{fig:classification}
\end{figure}

In case false alerts are acceptable in exchange for a higher detection rate, \name{}' sensitivity can be adapted, which is visualized in Figure~\ref{fig:classification}.
(Note that the ``benchmark'' graph will be described later.)
The figure shows the recall (i.e., ratio of detected evasions to all evasions) and precision (i.e., ratio of true alerts to all alerts) with respect to the decision threshold, along with the common classifier evaluation metrics F1 score and Matthew's correlation coefficient (MCC)~\cite{chicco2020matthews} (both semi-transparent).
Note that \name{} automatically selects a sensible threshold range and maps it to $[0,1]$ to ease configuration for non-expert users (cf. Appendix~\ref{sec:experiments-description}).

We can see in Figure~\ref{fig:classification} that the recall slowly declines when the threshold increases whereas the precision sharply increases around $0.5$.
The latter is caused by the extreme imbalance of the dataset because a slightly increasing false positive rate quickly leads to a large number of false alerts (cf. Appendix~\ref{sec:false-negatives} for an extended discussion).
Therefore, most users will probably choose thresholds near the default value to avoid significant numbers of false alerts without sacrificing too many detected evasions.
In practice, since an organization does not know the true number of attempted evasions, they should therefore set the threshold to a value that produces an acceptable number of false positives with respect to their analysts' resources.
For example, a detection rate of 75\,\% could be achieved on our dataset by selecting a lower threshold so that approximately one in ten alerts is a true evasion.
In this context, please also note that we would end up with significantly higher detection rates (at still few false alerts) if we had chosen a more balanced (but hence less realistic) dataset, since the precision curve would decline much slower.

Now that we quantified \name{}' classification performance, we will aim to improve the intuition on these results by comparing them with a benchmark approach.

\paragraph{Comparison with Benchmark Approach}

To the best of our knowledge, there is no related work on SIEM rule evasion detection, thus we cannot directly compare \name{} to related methods (cf. Section~\ref{sec:relatedwork}).
However, since we created all of our evasions from formerly detected attack events (i.e., \emph{matches}) in the course of our Sigma rule analysis (cf. Section~\ref{sec:analysis}), we can compare \name{} with a supervised classifier that learns from \emph{these matches} versus benign events (whereas \name{} learns from \emph{SIEM rules} versus benign events).
The rationale behind this comparison is to find out if SIEM rules might be missing information that is contained in the actual attack events, which would result in an inferior detection performance of \name{} as compared to this benchmark approach.

To perform this comparison, we repeated \name{}' training but instead of SIEM rules we used our matches as malicious samples.
As shown in Figure~\ref{fig:classification}, learning from these matches gave us similar results to those of \name{}.
At the default threshold, which corresponds to the highest possible recall without false alerts, both approaches perform almost identically (70\,\% detection rate), indicating that the information contained in SIEM rules is equally suited for evasion detection as the information contained in malicious events triggering these rules.
However, we deem the benchmark approach to be unrealistic in practice because it requires to manually create attack events for each SIEM rule, which is a tedious task and thus unlikely to be carried out whenever SIEM rules change (cf.\ Section~\ref{sec:amd}).
We are thus convinced that \name{} is a better choice for practical application in enterprise networks.

Next, we go beyond mere binary classification and assess \name{}' rule attribution component.
For this purpose, we evaluate whether \name{} succeeds in identifying the evaded SIEM rules when provided with evasions.

\subsection{Rule Attribution Helps Making Sense of Evasion Alerts (RQ2)}

\begin{figure}
  \includegraphics[width=\linewidth]{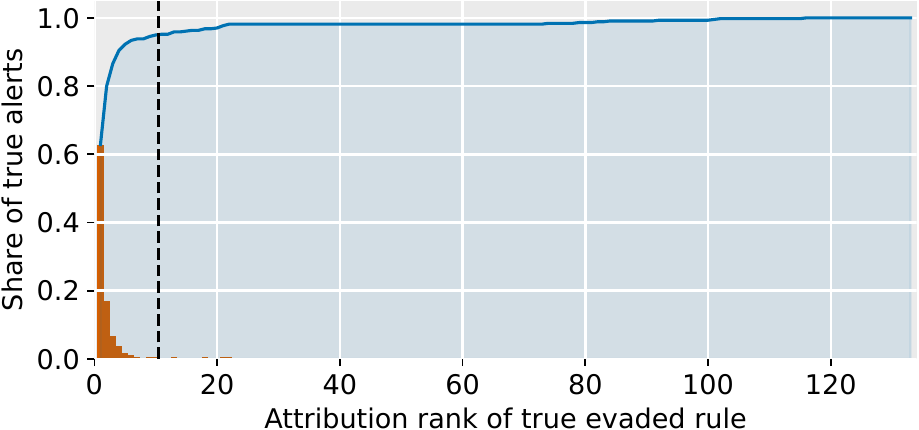}
  \caption{For each evasion alert, \name{} outputs a ranked list of likely evaded SIEM rules (rule attribution). Evaluated with our manually created evasions, the truly evaded rule is within the top 10 propositions for 95\,\% of the evasions.}
  \label{fig:attribution}
\end{figure}

Rule attribution is a unique feature of our adaptive misuse detection approach.
For each potential evasion (as detected by the misuse classification), \name{} suggests a ranked list of SIEM rules that were potentially evaded (cf. Section~\ref{sec:amd}).
Since we created at least three evasions for each evadable Sigma rule within the scope of our rule analysis, we can use these as ground truth, i.e., we know for each evasion which rule was evaded.
If the rule attribution worked perfectly, the evaded rule would be the highest-ranked for each evasion. %

We evaluated this expectation by applying the rule attribution to all detected evasions from the misuse classification step.
For each evasion, we checked the resulting rank of the true evaded rule.
The results are visualized in Figure~\ref{fig:attribution}.
The bars show a stack plot of the rank of the true evaded rules.
Ideally, all of these rules would appear in the first position.
We can see that this is the case at least for a large fraction of the evasions (63\,\%).
The blue curve shows the accumulated results, allowing to check the share of true rules that are contained in the top $n$ results.
Particularly, a vast majority of the true rules (95\,\%) are contained within the top~10.
This result indicates that the rule attribution largely succeeds in helping analysts to find out which rules were evaded.

Finally, after studying \name{}' performance in terms of classification and rule attribution, it remains to show that operation in an enterprise network is indeed feasible.

\subsection{\name{} Is Fit for Application (RQ3)}
\label{sec:speed-and-robustness}

To show that \name{} is well-suited for operation in an enterprise network, we evaluate three further aspects.
First, we show that \name{} is computationally efficient enough to handle enterprise-level throughput on commodity hardware.
Secondly, we examine the influence of evasions inadvertently contained in benign training data, showing that detection performance degrades gracefully.
Thirdly, we show that \name{} and the general concept of adaptive misuse detection are not restricted to Windows process creations but are also applicable to other rule and event types.

\paragraph{Computational Performance}

To find out if \name{} is efficient enough for practical application, we measured the time required for training as well as the event throughput during operation.
Both measurements were taken on a commodity Linux server with 40 physical CPU cores (80 virtual), 384\,GiB RAM, and a single SSD drive.
The \emph{training} was performed as described in Section~\ref{sec:amides} and the beginning of this section, i.e., using two weeks of benign events (80\,940\,832 events) and the Sigma rules described in Section~\ref{sec:analysis}.
It took 42~minutes on average ($n=10$, $\mu=2506$\,s, $\sigma=112$\,s), which allows for daily (re)training if desired.
However, most organizations will probably not update their SIEM rules that often, so a weekly training (or even less) probably suffices in most cases.

To measure the throughput during \emph{operation}, we let \name{} process our \emph{whole} dataset (i.e., four weeks of benign process creation events and 512 evasions, yielding 155\,373\,539 events in total).
\name{} required $0.0763$\,ms per event on average ($\sigma=3.19$\,ms) \emph{on a single core}, corresponding to 13\,102~events per second.
The overall processing took 994\,s on average ($n=10$, $\sigma=16.4$\,s) \emph{on all cores}, including reading the events from disk and storing them back to disk after adding the classification results and metrics.
The overall throughput on this server is thus 156\,311 events per second.
The peak number of (process creation) events per second in our dataset is 473, therefore \name{} is 330 times faster than required for live operation.
Much larger networks might require scaling up to multiple servers, which is straightforward because both training and operation can be parallelized.

\paragraph{Influence of Tainted Training Data}

Next, we consider the case that supposedly benign training data inadvertently contain evasions, which might arise when data from a productive network are used for training without careful review.
The purpose of this experiment is to find out how much detection performance degrades in this case, thereby verifying that \name{} is still useful if some missed evasions end up in the training data.
Since we could not find any reliable data on the prevalence of SIEM rule evasions in the wild, we added three different fractions (10-30\,\%) of our evasions to the \emph{benign} training set (and accordingly removed them from the validation set).
The largest fraction (30\,\%) represents a scenario where \name{} is already running during the data collection period, detecting 70\,\% of evasions (cf. Section~\ref{sec:classification-performance}) and no manual review is performed \emph{at all}, thus leaving 30\,\% of the attempted evasions in the benign training data.

\begin{figure}
  \includegraphics[width=\linewidth]{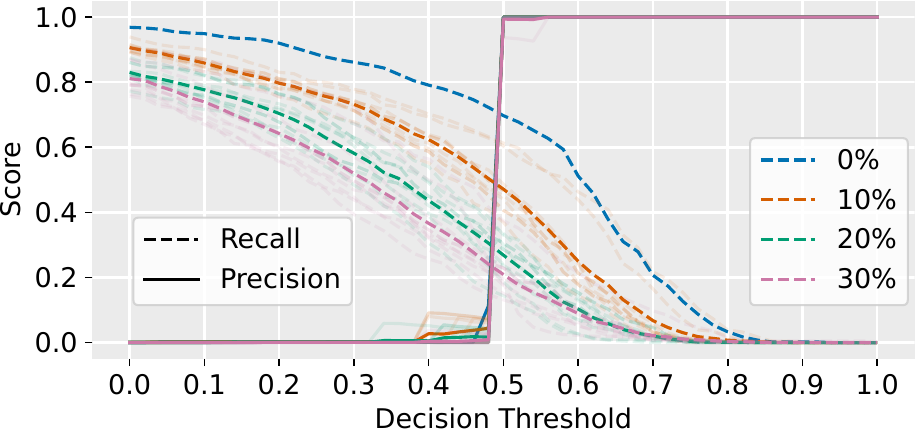}
  \caption{We added different fractions of our evasions to the benign training data to simulate heavy tainting with undetected evasions. As expected, classification performance significantly declines but still many evasions are detected.}
  \label{fig:tainted}
\end{figure}

We performed training and validation in the same way as described before, executing 10 runs per fraction with pseudo-randomly sampled evasions.
The results are depicted in Figure~\ref{fig:tainted}, showing all runs as semi-transparent lines and their respective means as opaque lines.
As expected, since \name{} is based on two-class supervised learning, detection performance decreases with an increasing fraction of tainted data.
These measurements support our intuition that it is indeed advisable to manually check the benign data for potential evasions prior to training.
However, \name{} still detects many evasions without false alerts, thus succeeding to significantly reduce detection blind spots even under heavily tainted data.

\begin{table*}[!b]
  \caption{The Sigma ruleset contains evadable rules for other event types than process creation. Three examples are shown here.}
  \small
  \centering
  \begin{tabularx}{\textwidth}{llll}
    \toprule
    \textbf{Event type} & \textbf{Sample affected rule} & \textbf{Affected search term} & \textbf{Sample evasion} \\
    \midrule
    Web request & proxy\_ios\_implant & */list/suc?name=* & http://.../list//suc?name=foo \\
    Registry change & susp\_run\_key\_img\_folder & *C:\textbackslash{}Windows\textbackslash{}Temp\textbackslash{}* & \%\%windir\%\%\textbackslash{}Temp\textbackslash{}foo.exe \\
    PowerShell & clear\_powershell\_history & ... -HistorySaveStyle SaveNothing & \$foo="SaveNothing"; ... -HistorySaveStyle \$foo \\
    \bottomrule
  \end{tabularx}
  \label{tab:newevasions}
\end{table*}

\begin{table*}[!b]
  \caption{We prototypically evaluated \name{} on the three additional rule/event types shown here to assess its generalizability.}
  \small
  \centering
  \begin{tabularx}{\textwidth}{llll}
    \toprule
    \textbf{Event type} & \textbf{Event provider and IDs} & \textbf{Event fields} & \textbf{Sigma rule fields} \\
    \midrule
    Web request & All Packetbeat events containing url.full & url.full & c-uri*, cs-uri*, cs-host, r-dns \\
    Registry change & Microsoft-Windows-Sysmon IDs 12-14 & winlog.event\_data.TargetObject, ...Details & TargetObject, Details \\
    PowerShell & Microsoft-Windows-PowerShell ID 4104 & powershell.file.script\_block\_text & ScriptBlockText, Message \\
    \bottomrule
  \end{tabularx}
  \label{tab:newtypes}
\end{table*}

\paragraph{Applicability to Other Rule and Event Types}
\label{sec:rule-types}

Our quantitative analysis of SIEM rules for evasions as well as our previous evaluation steps are based on one type of Sigma rules, namely, those acting on Windows process creation events (as argued in Section~\ref{sec:analysis}).
While this rule/event type is the most prevalent in the Sigma ruleset (292 out of 707 rules at the time of analysis) and highly prone to evasion (cf. Section~\ref{sec:analysis}), we also looked through all other Sigma rules and found that at least some of them can be evaded in similar ways.
Due to the high effort of manual rule analysis, we could not analyze all of them in detail to achieve statistically significant results, but still succeeded to find three prototypical evasions for rules acting on three fundamentally different event types, respectively: (1) web requests (47 rules in total), (2) Windows registry changes (40 rules), and (3) PowerShell script executions (25 rules).
Examples are given in Table~\ref{tab:newevasions}.

\begin{figure}
  \centering
  \includegraphics[width=\linewidth]{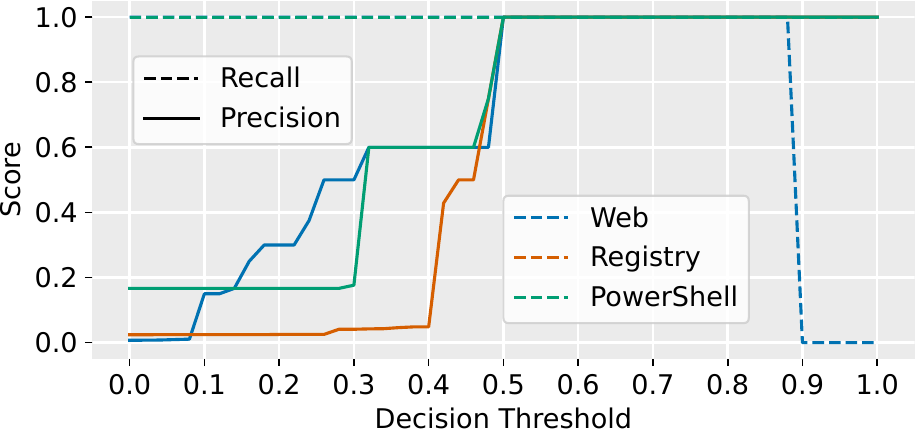}
  \caption{\name{} correctly classifies three additional datasets (each comprising synthetic benign events from a testbed and three prototypical evasions), indicating its applicability to other rule and event types than Windows process creation.}
  \label{fig:classification-newtypes}
\end{figure}

To demonstrate that our concept of adaptive misuse detection and particularly \name{} are applicable to other rule and event types than process creation, we evaluated \name{} against each of the three aforementioned types.
For this purpose, we utilized synthetic benign events from the testbed SOCBED~\cite{uetz2021socbed} since we did not have access to events of these types from the enterprise network.
Training and validation were performed in exactly the same way as for the process creation rules/events, only exchanging the utilized rule and event fields with respect to the event type (see Table~\ref{tab:newtypes} for details).
\name{} achieved perfect misuse classification performance (i.e., no missed evasions or false alerts) on all three types, as shown in Figure~\ref{fig:classification-newtypes}.
Furthermore, the rule attribution correctly ranked the evaded rules highest for each type.
While these results are not statistically significant due to the small number of evasions and the synthetically generated benign events (having limited quantity and variability), they still indicate that adaptive misuse detection and particularly \name{} are applicable to diverse rule and event types.

Overall, our evaluation has illustrated the feasibility of learning from SIEM rules versus benign events for the purpose of SIEM rule evasion detection, as proposed by our concept of adaptive misuse detection.
We have shown that the benefits remain even under real-world constraints involving limited analyst resources, high event volumes, and tainted training data.
Our proof-of-concept implementation, \name{}, therefore significantly reduces detection blind spots as revealed by our analysis of widespread SIEM rules. %

\section{Discussion and Limitations}
\label{sec:discussion}

In the following, we discuss the overall impact of our work as well as limitations of our concept and implementation.

To begin with, despite analyzing almost 300 widespread SIEM rules in detail, our work cannot draw a comprehensive picture of the detection blind spots arising from evadable rules in enterprise networks.
\textbf{Further work is needed to analyze additional sources and types of SIEM rules.}
Still, we showed that evasions pose a significant threat and organizations should strive to detect them, e.g., by using \name{}.

Next, we reflect on our main idea to train a classifier on SIEM rules versus benign events.
\textbf{Inherently, our approach can only detect evasions if (1) they manifest themselves in SIEM events and (2) these events are more similar to rule signatures than to benign events} (with respect to the extracted features).
Therefore, our approach cannot detect fundamentally different classes of evasion attacks such as (undetected) code injection into benign processes~\cite{barabosch2018injection}.

Furthermore, since we deal with an adversarial classification task here, \textbf{attackers could try to evade \name{} itself}, which is theoretically possible in two ways.
First, by tainting the training data though a large number of evasions over a long time.
We deem this to be unrealistic because \name{} would likely detect these evasions and they could thus be excluded from the training set.
Only when training for the first time, care should be taken to avoid a significant number of evasions in the benign data (cf. Section~\ref{sec:speed-and-robustness}).
Secondly, attackers could try to craft evasions that get classified as benign, e.g., by adding terms to malicious command lines that appear frequently in benign events but not in rules, while preserving a functioning command.
While we deem this attack to be difficult in practice because attackers only have one shot to avoid detection, it should be studied in future work to account for the ongoing rat race between attackers and defenders.

From a similar perspective, \textbf{our approach is not meant to detect completely novel attack types} but otherwise undetected instances of attacks with already existing detection rules.
To detect unknown attack types as well, adaptive misuse detection could be complemented with anomaly detection.
However, the latter may produce large numbers of false alerts and its ability to detect unknown attacks is arguable~\cite{gates2006challenging,sommer2010outside}.
Hence, organizations should consider and evaluate whether additional anomaly detection is beneficial to them.

Shifting the focus away from the basic concept, \textbf{some aspects of our implementation leave room for future improvements}.
In this work, we focus on Windows process creation events, particularly their command line field.
While we prototypically evaluated three additional event types, future work could assess these and others in more detail to further reduce potential detection blind spots.
Moreover, \name{} does not yet support the full Sigma specification.
For example, it correctly handles \texttt{NOT} statements within signatures, but does not yet support other statements such as \texttt{1 of them}.

Finally, we discuss the cost-benefit ratio for organizations extending a conventional misuse detection system with \name{}.
On the cost side, there is an initial effort to (1)~install \name{}, (2)~establish an (automated) training process, and (3)~manually examine the initial training data for potential evasions.
Additionally, ongoing effort arises from sifting the raised evasion alerts (which, according to our evaluation, contain very few false positives).
On the benefit side, \name{} will likely detect a majority of SIEM rule evasions, which would otherwise go undetected.
Thus, \name{} reduces detection blind spots and enables a timely response to such cyberattacks.
While there seem to be no reliable data on the prevalence of such evasions, \textbf{we expect the benefits to outweigh the costs of \name{} for many organizations} in view of the significantly reduced risk of undetected cyberattacks.

\section{Related Work}
\label{sec:relatedwork}

The main goals of our work are to \emph{identify} and subsequently \emph{reduce} detection blind spots in enterprise networks that arise from SIEM rule evasions.
To the best of our knowledge, these issues have not been addressed by other work before, hence, conducting an empirical comparison with related work is not possible.
Still, there are numerous works that share the overarching goal of threat detection in enterprise networks.
We begin by discussing the most similar ideas, namely, approaches utilizing events of any kind (but not necessarily from a SIEM system) to detect evasions in a broad sense.
Afterwards, we discuss approaches that also utilize events but do not address evasions.
Lastly, vice versa, we discuss approaches related to evasion detection but utilizing other data sources than events.

\paragraph{Event-Based Evasion Detection}

Ongun et al.~\cite{ongun2021lol} address the challenge of detecting adversarial executions of benign programs, which is a use case also covered by various of the analyzed Sigma rules.
The authors generate features from process creation command lines and train a multi-class classifier for five attack classes using an active learning approach.
In contrast to \name{}, attack classes must be defined manually and the active learning process requires ongoing analyst resources, whereas our approach leverages the ever-growing rule corpus of the Sigma community and therefore does not require any manual labeling.
Buyukkayhan et al.~\cite{buyukkayhan2017lens} utilize a combination of event records, static program analysis, and contextual information to discover programs behaving similarly to known malware as well as programs impersonating known benign programs.
While this approach can be seen as a type of evasion detection, it clearly differs from our goal to detect SIEM rule evasions.
Dreger et al.~\cite{dreger2005enhancing} utilize web server logs to detect evasions of a network-based intrusion detection system (NIDS) by correlating host-based and network-based events.
While this approach seems to work well for its specific use case, it cannot detect SIEM rule evasions.

\paragraph{Other Event-Based Threat Detection}

Numerous threat detection approaches utilize event records (just as \name{}) but pursue other goals than finding (SIEM) rule evasions.
Early works apply various learning methods to host-based audit logs for detecting anomalies or deriving detection rules~\cite{denning1987intrusion, endler1998intrusion, lee1999datamining}.
In contrast, publications within the last ten years often consider events from a large number of systems in a network or from multiple event sources to find anomalies~\cite{yen2013beehive, veeramachaneni2016ai2, pei2016hercule, roundy2017smoke, ho2021hopper}.
Particularly in the last few years, various works employ provenance graphs to model causal relations in system behavior, predominantly for finding suspicious activity~\cite{li2021provsurvey, hassan2020tactical, alsaheel2021atlas, yu2021alchemist}.
While all of these approaches address threat detection from events, they do not provide organizations with a straightforward way to reduce detection blind spots caused by SIEM detection rule evasions.

\paragraph{Other Evasion Detection}

The challenge of evasion detection has been studied for other data sources than event records as well, namely, network traffic and files (especially potential malware files).
Considering network traffic, Sohi et al.~\cite{sohi2021rnnids} enhance the detection performance of an NIDS by generating signature mutations using a neural network.
Varghese et al.~\cite{varghese2006evasion} split NIDS signatures into smaller pieces to counter evasions by TCP fragmentation.
Cheng et al.~\cite{cheng2012evasion} survey NIDS evasion techniques and assess their effectiveness.
Considering files, Shiel and O'Shaughnessy~\cite{shiel2019fuzzy} as well as Li et al.~\cite{li2015fuzzy} analyze fuzzy hashing to detect variants of known malware files.
Kapravelos et al.~\cite{kapravelos2013revolver} detect evasive behavior in malicious JavaScript files by comparing them to known malicious variants.
Aslan and Samet~\cite{aslan2020malware} survey further malware detection approaches.
Due to their fundamentally different data sources, the above approaches are not meant to detect SIEM rule evasions and are thus complementary to \name{}.

\section{Conclusion}
\label{sec:conclusion}

This work addresses evasions of SIEM detection rules, which can lead to critical detection blind spots within enterprise networks.
To quantify the risk of such evasions, we analyzed a set of widespread open-source SIEM rules, namely, Windows process creation rules from the official Sigma repository.
We found that out of 292 analyzed rules, 110 can by fully evaded and 19 can be partially evaded, potentially enabling adversaries to perform a large number of known attacks without being detected.
To remedy this situation, we proposed the novel concept of adaptive misuse detection, an extension of conventional (i.e., rule-based) misuse detection where incoming events are compared to SIEM rules on the one hand and known-benign events on the other hand to detect evasions and additionally estimate which rules were evaded.
Notably, this approach is well-suited for implementation in enterprise networks because it relies on already existing data (i.e., a SIEM ruleset and a set of historical benign events).

Using our open-source proof-of-concept implementation \name{}, we evaluated our approach on four weeks of benign process creation events from a large enterprise network and 512 manually created evasions.
At its default sensitivity, \name{} detected 70\,\% of the (otherwise undetected) evasions with zero false alerts.
\name{} processes \textasciitilde 156\,000 events per second in operation and requires 42~minutes for training on a commodity server with our dataset, making it fast enough for operation even in very large enterprise networks.
We also evaluated \name{}' capability to attribute evasions to evaded SIEM rule(s) and found that for 95\,\% of our evasions, the actually evaded rule was within \name{}' top 10 propositions, thus likely helping analysts in verifying and analyzing alerts.
Furthermore, we showed that \name{} still detects a significant share of evasions with no false alerts even if the benign training data are heavily tainted with evasions.

Future work should further examine widespread SIEM rules for potential evasions and expand the concept and implementation of adaptive misuse detection by additional event types and fields to increase the coverage of potential detection blind spots.
Overall, we showed that organizations relying on SIEM rules for cyberattack detection are prone to critical detection blind spots but can significantly reduce them with moderate effort by utilizing adaptive misuse detection.
Thus, our work contributes to an improved detection of cyberattacks in enterprise networks and ultimately to reducing the damage arising from data theft, sabotage, and extortion.

\section*{Acknowledgments}

We would like to express our gratitude to the SOC managers of the enterprise in which we performed our experiments for greatly supporting our ideas and work.
Moreover, we sincerely thank our shepherd and the reviewers for their comprehensive, nuanced, and encouraging comments and suggestions.

\begin{appendix}

\section{Implementation Details and Discussion}
\label{sec:experiments-description}

This section extends our description of \name{}' implementation in Section~\ref{sec:amides}, justifies our choices, and discusses potential alternatives.
Generally speaking, we see our main contribution in proving that our concept of adaptive misuse detection (particularly: learning from SIEM rules) succeeds in detecting evasions in enterprise networks, \emph{despite} using straightforward machine learning techniques.
Hence, we did not strive to (over-)optimize our results by tailoring the feature extraction to our specific evasion instances or comparing various classifiers and choosing the best one for our specific dataset.
Yet, we considered and/or tested several alternatives that seemed sensible, which are discussed in the following.

\paragraph{Feature Extraction}
\label{sec:feature-extraction}

We designed \name{}' feature extraction based on insights from creating evasions and sifting benign events.
In the filtering step (cf. Figure~\ref{fig:amides}), we remove characters from the command line string that adversaries can insert almost anywhere (even within words) to evade signatures without invalidating the command.
(We also use this technique for many of our evasions.)
While the exact character set for filtering is arguable, we chose characters known to allow for \texttt{cmd.exe} and PowerShell evasions~\cite{bohannon2017revoke,bohannon2019dosfuscation}.

The subsequent splitting step tokenizes the command line into terms.
We experimented with different splitting character sets optimized for command lines but eventually opted for a simple ``word'' splitting (regex \texttt{\textbackslash W}) since we discovered too many edge cases where other splittings lead to undesirable results.
Though removing some context information, we found word splitting to be the most robust and generalizable.

While sifting the enterprise events, we found a large number of command lines containing unique arguments such as timestamps or IDs, bloating the feature space without contributing relevant information.
Hence, we implemented the two pruning rules described in Section~\ref{sec:amides} to remove the majority of such irrelevant arguments.
We suppose that these rules fit other enterprise environments as well, but suggest to ensure for any new environment that they succeed in removing irrelevant terms while preserving the relevant ones.

Finally, the remaining terms are transformed into a tf-idf vector.
Since we have an extremely imbalanced dataset, we found it desirable to assign lower weights to frequent terms (as tf-idf does) to counter potential benign terms within an evasion dominating the classification decision.
We also tried regular and normalized bag-of-words vectorization, but got significantly worse results (as expected).
So far, we did not experiment with context-sensitive methods such as word2vec~\cite{mikolov2013distributed} because the utilized Sigma signatures are often very short, consisting of only one or few words, whereas the benign events comprise full command lines.
Together with the fact that argument positions are often exchangeable, we do not expect context-sensitive methods to perform well for our approach, hence leaving them to potential future work.

\paragraph{Classification}

Various classifiers come into question for intrusion detection tasks~\cite{panigrahi2021performance,kus2022false}.
Since a sound comparison would go beyond the scope of this work, we opted for linear SVMs according to a best-practice guide~\cite{chooseestimator}.
We found them to be suitable for our data both in terms of classification and computational performance.
Moreover, linear SVMs enable straightforward analysis of the influence of input terms on the classification result, facilitating effective feature design.

For hyperparameter tuning, we perform grid search over 25 exponentially growing C values between $0.01$ and $10$, with class balancing (\texttt{class\_weight} in scitkit-learn) on and off, respectively.
We also tested other SVM kernels (RBF, polynomial, and sigmoid) but did not get significantly better results.
Overall, we cannot rule out that other classifiers might yield better results, but leave their evaluation to future work.

\paragraph{Threshold Configuration}

The ideal decision threshold in operation usually differs from the one resulting from training, mainly because security analysts in enterprises need to investigate alerts from multiple security systems within limited time.
Hence, both too many and too few alerts can lead to missed attacks (cf. Section~\ref{sec:background}).
Therefore, we decided to set the default threshold for operation such that \name{} raises a configurable number $N$ of false alerts \emph{per day on average}, thus matching available analyst resources.
This feature is implemented simply by setting the threshold such that $ND$ false positives occur in the training set, where $D$ is the length of the training set in days.
This method assumes similar statistical properties of benign events in training and operation, which we deem justified when performing regular retraining.
For our evaluation, we chose $N=0$ for the sake of simplicity, effectively placing the default decision threshold just above the worst-classified benign training sample (cf. Appendix~\ref{sec:false-negatives}).

Furthermore, besides a sensible \emph{default} threshold, we also calculate a sensible threshold \emph{range} with the goal of providing non-expert users with a simple configuration parameter between zero and one, where $0.5$ is the default threshold, allowing to put a strong emphasis on either recall or precision.
We implement this by linearly mapping a subinterval of the SVM decision function values to $[0,1]$ such that the default threshold lies at $0.5$ and the MCC curve remains above $0.1$ over most of the interval (note that MCC is more robust than F1 score or individual precision/recall limits~\cite{chicco2020matthews}).

\section{Analysis of Classification Performance}
\label{sec:false-negatives}

\begin{figure}
  \centering
  \includegraphics[width=\linewidth]{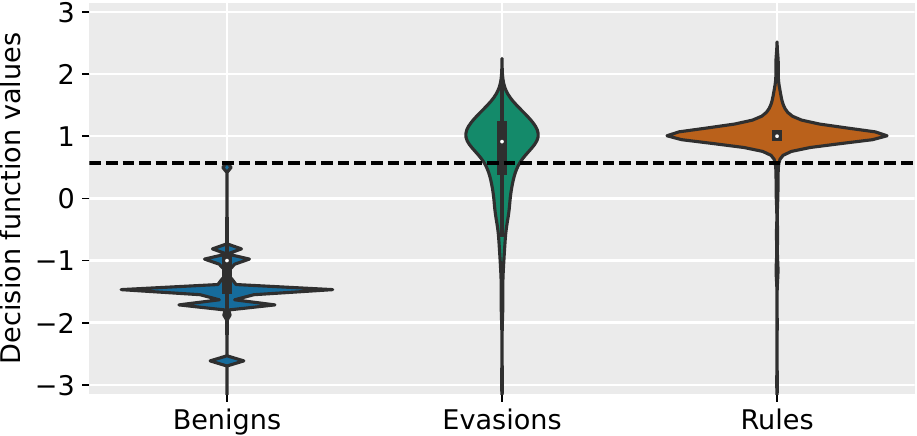}
  \caption{The imperfect class separability and high imbalance would result in a \textasciitilde 70\,\% evasion detection rate in practice.}
  \label{fig:distribution}
\end{figure}

To assess \name{}' misuse classification performance, we statistically analyzed its output for all benign events, evasion events, and Sigma rules.
Figure~\ref{fig:distribution} shows the distributions of the resulting SVM decision function values.
The dashed line depicts the default threshold as described in Appendix~\ref{sec:feature-extraction} for $N=0$, which also yields zero false alerts on the validation set (which is not generally the case since there could be worse-classified benign events than in the training set).
We can see that the classifier largely succeeds in separating the training classes (i.e., benigns and rules), with the median samples being located approximately at $-1$ and $1$, respectively, but no perfect separation is achieved.
Furthermore, the plot reveals an uneven distribution of the benign events, which we can attribute to large clusters of similar command lines.

Looking at the evasions, both median and average values are slightly below those of the rules.
Particularly, we can see that a significant fraction (30\,\%) exhibits smaller values than the highest-valued benign cluster just below the default threshold.
Even though this cluster comprises only a small fraction of the benign events, the extreme class imbalance (factor \textasciitilde 145\,000) causes a sharp decline in precision (cf. Figure~\ref{fig:classification}), rendering lower thresholds impractical due to many false alerts, thus yielding the reported 70\,\% detection rate.

We also analyzed the contributions of all features to the classification results.
We found that many evasions contain terms that appear in benign events but not in rules, thus explaining their lower decision function values.
Particularly, signatures often omit less relevant parts of malicious command lines (e.g., the full path of an executable), hence such terms in evasions influence their classification towards benignity.
For example, the term \texttt{system32} has a strong weight towards benignity since it frequently appears in benign events but rarely in rules.
Vice versa, there is a cluster of \textasciitilde 188\,k misclassified benign executions of \texttt{winlogon.exe} close to the default threshold because this binary appears in several rules (with additional conditions that \name{} currently ignores due to our restriction to the command line field).
Consequently, a straightforward approach to improve \name{}' detection performance in practice would be to filter its alerts by a list of known false positives (i.e., misclassified benign command lines), allowing to set a lower threshold and thus detect more evasions without increasing the number of false alerts.

Overall, while this analysis reveals potential limitations of learning from SIEM rules, we nonetheless consider our concept of adaptive misuse detection (and \name{} in particular) highly beneficial in practice since it facilitates a significant reduction of detection blind spots in enterprise networks.


\section{Artifact Appendix}

\subsection{Abstract}
\label{sec:artifact-abstract}

To facilitate reproduction of our experiments as well as extensions and improvements of \name{}, our artifact comprises five components: (1)~the \name{} source code along with automation scripts, (2)~the set of SIEM rules from the public Sigma repository that we analyzed for possible evasions, (3)~the set of \emph{matches} that we created, i.e., SIEM events triggering the aforementioned detection rules, (4)~the set of \emph{evasions} that we created, i.e., \emph{matches} adapted such that the executed commands achieve the exact same goal without triggering the respective rule, and (5)~a set of \emph{benign} SIEM events that we created using the open-source testbed SOCBED (since we are not allowed to share events from the real enterprise network).
Together, these components enable reproduction of our experiments and thus confirmation of our claims.

\subsection{Description \& Requirements}

\subsubsection{Security, privacy, and ethical concerns}
\label{sec:artifact-ethical}

The artifact does not pose a security risk when downloaded and executed since none of the potentially malicious command lines within the SIEM events and rules are run during the experiments.
However, security software such as endpoint protection products might still raise alerts due to these command lines being contained in the event and/or rule files.

Due to ethical concerns, we will not make our full set of evasions publicly available (as stated in the paper).
However, our artifact contains a small number of evasions for testing purposes, namely, those given as examples in the paper in Tables 1 and 3.
Please contact us if you require the full set of evasions for your research.

\subsubsection{How to access}

The artifact is available on GitHub\footnote{\url{https://github.com/fkie-cad/amides/releases/tag/v1.0.0}}.
Please start by cloning or downloading the repository on a commodity computer running Linux or macOS.

\subsubsection{Hardware dependencies}

Assuming small training and validation datasets such as those provided with the artifact, \name{} runs on a commodity computer with a minimum of 8\,GB of RAM and requires around 2\,GB of disk space.
For larger training and validation datasets, more RAM and disk space are required.
A fast CPU has a positive impact on the duration of training and validation.

\subsubsection{Software dependencies}

\name{} is written in Python. 
The repository contains a list of Python package requirements that need to be installed in order to use \name{}.
All of the requirements can be installed from PyPI using pip.
For convenience, the repository contains a Dockerfile and automation scripts to build containers running \name{} and reproduce our experiments.
Building and operating the containers requires a Docker installation. 
The containers should run on any operating system, however, our automation scripts are currently written for Linux and macOS.
Please refer to the README file for further information.

\subsubsection{Benchmarks}
\label{sec:artifact-benchmarks}

The majority of experiments in our paper (all except ``Applicability to Other Rule and Event Types'') are based on real benign SIEM events from a large enterprise network that are strictly prohibited to be taken off the premises.
However, as stated in §6 ``Datasets and Ethical Considerations'', we additionally created synthetic benign events using the open-source testbed \emph{SOCBED} (which were also used for the aforementioned experiment) to facilitate reproduction of our experiments and confirmation of our claims.
These synthetic benign events, along with all other data required for reproducing our experiments (cf. §\ref{sec:artifact-abstract}) are contained in the repository except for the full set of evasions (cf. §\ref{sec:artifact-ethical}).

\subsection{Set-up}

\subsubsection{Installation}
\label{sec:artifact-installation}

After cloning or downloading the repository, please execute the steps described in the ``Building the Quickstart Environment'' section of the README file to build the image and containers that will run \name{} and reproduce our experiments.
Alternatively, \name{} can be installed locally, which is described in the ``Installation'' section of the README file.

\subsubsection{Basic Test}

Running the Installation (§\ref{sec:artifact-installation}) and Experiments (§\ref{sec:artifact-experiments}) instructions without errors ensures that the quickstart environment and \name{} are functioning properly.
If any of the automated steps fail, corresponding output messages will be generated. 
In case \name{} was installed locally, the successful execution of its unit tests indicates that all components are functioning.
Please refer to the ``Testing'' section of the README file for instructions on how to execute unit tests.

\subsection{Evaluation workflow}

\subsubsection{Major Claims}

\begin{compactdesc}

    \item[(C1):]
    \name{} detects a majority of our crafted evasions without any false alerts (cf. §6.1 ``Classification Performance'' and Figure~3, plot ``\name{}'').
    
    \item[(C2):]
    \name{}' classification performance keeps up with a (much more costly) benchmark approach (cf. §6.1 ``Comparison with Benchmark Approach'' and Figure~3).
    
    \item[(C3):]
    \name{} helps security analysts to attribute its evasion alerts to potentially evaded SIEM rules (cf. §6.2 and Figure~4).
    
    \item[(C4):]
    \name{} degrades gracefully in detection performance when the training set is tainted with attacks (cf. §6.3 ``Influence of Tainted Training Data'' and Figure~5).
    
    \item[(C5):]
    \name{} is applicable to multiple SIEM rule and event types (cf. §6.3 ``Applicability to Other Rule and Event Types'' and Figure~6).
    
\end{compactdesc}

\subsubsection{Experiments}
\label{sec:artifact-experiments}

Please follow the instructions given in the ``Running Experiments using the Quickstart Environment'' section of the README file.
The corresponding container will then automatically execute all four experiments (E1-E4, see below), reproduce the above-mentioned plots as PDF files, and place them in the specified folder.
Note that the file names of the generated plots include the major claims they are addressing (i.e., C1-C5).
In case \name{} was installed locally, please refer to the ``Running Experiments'' section of the README file, which also contains more details on the experiments.

Since the benign SIEM events in the repository are not from a real enterprise network but generated by a testbed (cf. §\ref{sec:artifact-benchmarks}), the plots corresponding to Figures~3, 4, and 5 will look different compared to the paper.
More precisely, the results on the testbed-generated data are significantly better because the number of benign events is much lower compared to the real enterprise events, leading to an easier classification task for \name{} (cf. §6.1).
Still, these results confirm our claims and facilitate future research.

The evaluation comprises the four subsequently mentioned experiments and requires approximately 45~human-minutes, 20~compute-minutes, and around 2\,GB of disk space.
As a congruence check, we provide the correct output in the document \texttt{Paper Supplement.pdf} in the repository.

\vspace{2mm}
\begin{compactdesc}

    \item[(E1):]
    \textit{Classification Performance} [2~compute-minutes]:
    \name{}' evasion detection performance is compared to a benchmark approach that was trained using matches instead of SIEM rules.
    Result: \texttt{figure\_3\_c1\_c2\_misuse\_classification.pdf} in the \texttt{amides/plots/process\_creation} folder.
    
    \item[(E2):]
    \textit{Rule Attribution} [2~compute-minutes]:
    \name{}' rule attribution performance is evaluated by assessing if detected evasions are correctly assigned to the corresponding Sigma detection rules.
    Result: \texttt{figure\_4\_c3\_rule\_attribution.pdf} in the \texttt{amides/plots/process\_creation} folder.
    
    \item[(E3):]
    \textit{Tainted Training Data} [10~compute-minutes]:
    \name{}' evasion detection performance is evaluated after training data has been tainted using evasions.
    The tainting is repeated for different fractions of events and different events for each fraction. 
    Result: \texttt{figure\_5\_c4\_tainted\_training.pdf} in the \texttt{amides/plots/process\_creation} folder.
    
    \item[(E4):]
    \textit{Other Rule and Event Types} [2 compute-minutes]:
    \name{}' evasion detection performance is evaluated for three additional rule and event types (Windows PowerShell, Windows Registry, and Web).
    Result: \texttt{figure\_6\_c5\_classification\_other\_types.pdf} in the \texttt{amides/plots} folder.
    
\end{compactdesc}

\subsection{Notes on Reusability}
\label{sec:reuse}

\name{} is fit for application in enterprise networks.
To this end, users can perform the required training with their own data (i.e., a SIEM ruleset and a set of up-to-date benign SIEM events).
The resulting model can then be loaded and applied to SIEM events by the open-source log data processor \emph{Logprep}\footnote{\url{https://github.com/fkie-cad/logprep}}, for which we implemented and published an \name{} processor\footnote{\url{https://logprep.readthedocs.io/en/latest/user_manual/configuration/processor.html\#amides}}.
For more information on how to create models for \name{} from scratch, refer to the ``Running Custom Experiments'' section of the README file.

\subsection{Version}

Based on the LaTeX template for Artifact Evaluation V20231005. Submission,
reviewing and badging methodology followed for the evaluation of this artifact
can be found at \url{https://secartifacts.github.io/usenixsec2024/}.

\end{appendix}

\end{document}